# Magnetization study of RuSr$_2$Y$_{1.5}$Ce$_{0.5}$Cu$_2$O$_{10}$ (Ru-1222)

I. Felner

Racah Institute of Physics, The Hebrew University, Jerusalem, 91904, Israel.

V.P.S. Awana[1,2] and E. Takayama-Muromachi[2]

[1]National Physical Laboratory, K.S. Krishnan Marg, New Delhi 110012 India
[2]Superconducting Materials Center, NIMS, 1-1 Namiki, Tsukuba, Ibaraki, 305 0044, Japan

We have studied the magnetic properties of the non-superconducting RuSr$_2$Ln$_{1.5}$Ce$_{0.5}$Cu$_2$O$_{10}$ (Ln=Y, Dy and Ho, Ru-1222) compounds synthesized under high pressure (6 Gpa) at elevated temperature. The materials become magnetically ordered at $T_M$ =152(2) K regardless of Ln. The wide ferromagnetic-like hysteresis loops which open at 5 K, close themselves around $T_{irr}$= 90-100 K and the remanent magnetizations ($M_{rem}$) and the coercive fields ($H_C$) become zero. Surprisingly, at $T_{irr}<T< T_M$ a reappearance of the $M_{rem}$ and $H_C$ (with a peak at 120-130 K) is observed for all three samples studied. For the non-magnetic Ln=Y compound, the extracted saturation moment at 5 K and the effective paramagnetic moment are is 0.75 and 2.05 $\mu_B$ /Ru, values which are close to the expected 1 $\mu_B$ and 1.73 $\mu_B$ respectively, for the low-spin state of Ru$^{5+}$. We argue that the Ru-1222 system becomes (i) anti-ferromagnetically (AFM) ordered at $T_M$. In this range a metamagnetic transition is induced by the applied field (ii). At $T_{irr}$ < $T_M$, weak-ferromagnetism (W-FM) is induced by the canting of the Ru moments.

## Introduction

The interplay of magnetism and superconductivity (SC) is a fundamental problem in condensed matter physics and it has been studied experimentally and theoretically for almost four decades. Coexistence of weak-ferromagnetism (W-FM) and SC was discovered a few years ago in RuSr$_2$R$_{2-x}$Ce$_x$Cu$_2$O$_{10}$ (R=Eu and Gd, Ru-1222) layered cuprate systems [1-2], and more recently[3] in RuSr$_2$GdCu$_2$O$_8$ (Ru-1212). The SC charge carriers originate from the CuO$_2$ planes and the W-FM state is confined to the Ru layers. In both systems, the magnetic order does not vanish when SC sets in at $T_C$, but remains unchanged and coexists with the SC state. The Ru-1222 materials (for

R=Eu and Gd) display a magnetic transition at $T_M$= 125-180 K and bulk SC below $T_C$ = 32-50 K ($T_M > T_c$) depending on oxygen concentration and sample preparation[1]. The hole doping of the Cu-O planes, which results in metallic behavior and SC, can be optimized with appropriate variation of the R/Ce ratio[4]. SC occurs for Ce contents of 0.4-0.8, and the highest $T_C$ was obtained for Ce=0.6. SC survives because the Ru magnetic moments probably align in the basal planes, which are practically decoupled from the $CuO_2$ planes, so that there is no pair breaking. X-ray-absorption spectroscopy (XAS)[5] taken at the K edge of Ru, at room temperature reveals that the Ru ions are basically $Ru^{5+}$ or $Ru^{4.75+}$( see Ref 6). Ru remains pentavalent irrespective of the Ce concentration, which means that there is no charge transfer to the Ru-O layers with increasing Ce concentration[7]. The remaining unresolved question is whether the $Ru^{5+}$ in Ru-1222 is in its low (S=1/2) or high (S=3/2) spin state. It is also apparent, that bulk SC only, appears in the iso-structural $MSr_2R_{2-x}Ce_xCu_2O_{10}$ (M= Nb and Ta) with $T_C$~28-30 K, in which the M ions are pentavalent [8].

The SC state in Ru-1222, is well established and understandable. Specific heat studies show a sizeable typical jump at $T_C$ and the magnitude of the ΔC/T (0.08 mJ/gK$^2$) indicates clearly the presence of bulk SC [9]. The specific heat anomaly is independent of the applied magnetic field. The temperature dependence of the magneto-resistance data[10] yield the upper critical field of $H_{c2}(0)$= 39 T and the coherence length of $\zeta(0)$ =140 Å along the $CuO_2$ planes. Due to the granular nature of the materials the critical current density at 5 K is extremely small; $J_c(0)$ =22 A/cm$^2$ as compared to other high $T_C$ superconducting materials[11]. Below $T_C$, the magneto-resistance ΔR(H) = R(H)-R(0) is positive and unexpected hysteresis loops are observed. ΔR(H) on decreasing the applied field (H) is much smaller than the ΔR(H) for increasing H. The width of the loops depends strongly on the weak-link properties. Similar hysteresis loops are observed in the SC, non-magnetic Nb-1222 ($T_C$ =28 K), thus excluding that the hysteresis phenomenon is caused by the coexistence of SC and magnetic states[11]. Scanning tunneling spectroscopy [1] and magneto-optic experiments [12] have demonstrated that all materials are microscopically uniform with no evidence for spatial phase separation of SC and magnetic regions. That is, both states coexist intrinsically on the microscopic scale. Studied of Zn substitution for Cu in oxygen annealed $RuSr_2Eu_{2-x}Ce_x(Cu_{1-x}Zn_x)_2O_{10}$ (x=0, 0.01 and 0.025) reveal, that Zn reduces $T_C$, from $T_C$ =38 K for x=0, to 26 K for 0.01 and that for x=0.025, the material is not SC down to 4.2 K. On the other hand, the magnetic state of the Ru sublattice is not effected by the presence or absence the SC State, indicating that the two states are practically decoupled[13].

One of the most disputed questions is the magnetic structure in Ru-1222. In contrast to the Ru-1212 system in which the antiferomagnetic (AFM) nature of the Ru sublattice has been determined by neutron diffraction studies [14-15], the detailed magnetic features of the Ru-1222 system are lacking. The accumulated results are compatible with two alternative scenarios, both of which are used in understanding the qualitative features at low applied fields. (A) Going from high to low temperatures, the magnetic behavior is basically divided into two regions [4]. (i) Depending on Ce content, at $T_M$, *all* the material becomes AFM ordered. (ii) At $T_{irr}$ ($<T_M$), a W-FM state is induced, which originates from canting of the Ru moments. This canting is probably a result of an anti-symmetric exchange coupling of the Dzyaloshinsky-Moriya (DM) type[16] between neighboring Ru moments, induced by the tilting of the $RuO_6$ octahedra from the crystallographic c axis [17]. This local distortion causes the adjacent spins to cant slightly out of their original AFM direction and to align a component of the moments with the direction of the applied field. At $T_C < T_{irr}$, SC is induced and both the SC and the WFM states coexist intrinsically on a microscopic scale. (B) Detailed analysis of the magnetization under various thermal-magnetic conditions suggests a *magnetic* phase separation of Ru-1222 into FM and AFM[18] nano-domain species inside the crystal grains. A minor part of the material becomes FM at $T_M$, whereas the major part orders AFM at a lower temperature and becomes SC at $T_C$. In this scenario, the unusual SC state is well understood.

In attempting to understand the W-FM state in Ru-1222, we report here a detailed magnetization study of $RuSr_2Y_{1.5}Ce_{0.5}Cu_2O_{10}$ (Ru-1222Y) which has been synthesized at 6GPa at 1200° C[19]. Non-magnetic Y has replaced the magnetic Eu and/or Gd ions, which permits an easier direct interpretation of the intrinsic Ru magnetism. The results will be compared to the data obtained for $RuSr_2Dy_{1.5}Ce_{0.5}Cu_2O_{10}$ (Ru-1222Dy) $RuSr_2Ho_{1.5}Ce_{0.5}Cu_2O_{10}$ (Ru1222Ho) both synthesized under the same conditions [19]. We show below that the magnetic structure of all samples studies can be interpreted only by assuming model A discussed above. None of the samples described here is superconducting. Attempts to induce SC by a annealing the materials under high oxygen pressure at elevated temperatures have failed.

**Experimental details**

Ceramic samples with nominal composition $RuSr_2Ln_{1.5}Ce_{0.5}Cu_2O_{10}$ (Ln=Y, Dy and Ho) under 6 GPa at 1200°C for 2 hours were prepared by a solid-state reaction technique as described in Ref.19. Determination of the absolute oxygen content in these materials is difficult because $CeO_2$ is not completely reducible to a stoichiometric oxide when heated to high temperatures. Powder X-ray diffraction (XRD) measurements confirmed the tetragonal structure (SG I4/mmm) and yield the lattice parameters: a=3.824(1), 3.819(1) and 3.813(1) Å and c=28.445, 28.439(1) and 28.419(1) Å for Ln= Dy, Y and Ho respectively. Ru-1222Dy and Ru-1222Ho are single-phase

materials whereas the Ru-1222Y pattern shows three small (less than 3%) un identified extra peaks[19]. Zero-field-cooled (ZFC), field-cooled (FC) dc magnetic measurements at various applied field in the range of 5-300 K were performed in a commercial (Quantum Design) super-conducting quantum interference device (SQUID) magnetometer. Ac susceptibility was measured (at $H_{dc}=0$) by a homemade probe, with excitation frequency and amplitude of 733 Hz and 30 mOe respectively, inserted in the SQUID magnetometer.

**Experimental results**

In contrast to the magneto-superconducting $RuSr_2Ln_{1.5}Ce_{0.5}Cu_2O_{10}$ (Ln =Eu and Gd), which are synthesized at ambient pressure, the heavy Ln (and Y) materials can be synthesized only under high pressures. The samples obtained are only magnetically ordered. Attempt to induce SC by annealing the Ru-1222Y material under 75 atm. oxygen at 800 °C for 6 hours, leads to decomposition into a $RuSr_2YO_6$ (1216) phase[20]. The ionic radius of $Dy^{3+}$ is 0.91 Å as compared to 0.90 Å for both $Y^{3+}$ and $Ho^{3+}$ ions. Therefore the unit cell volume (V) of $RuSr_2Dy_{1.5}Ce_{0.5}Cu_2O_{10}$ (415.9 Å$^3$), is a bit larger than that of Ln=Y and Ho 414.7 Å$^3$ and 413.2 Å$^3$ respectively.

The temperature dependence of the normalized real ac susceptibility curves for the $RuSr_2Ln_{1.5}Ce_{0.5}Cu_2O_{10}$ system are presented in Fig. 1. It is readily observed that none of the materials is SC down to 4.2 K. The two peaks observed, namely the major one at 91(1) 105 (1) and 107(1) K, and the minor one (Fig. 1 inset) at T=120(1), 128(1) and 130(1) K, for Ln= Dy, Y and Ho respectively, are both inversely proportional to V listed above. The increase of the signals below 40 K for Ln= Dy and Ho is related to their large paramagnetic (PM) contribution at low temperatures. None of the two peaks is the magnetic transition $T_M(Ru)$ of the system as discussed below. Beside these differences, the magnetic behavior of all three samples is similar and for the sake of brevity we'll describe in detail the magnetic properties of Ru-1222Y in which Y is not magnetic.

ZFC and FC dc magnetic measurements were performed over a broad range of applied magnetic fields and typical M/H curves for Ru-1222Y measured at 15 (and 500 Oe), are shown in Fig.2. The two curves merge twice: at $T_{irr}$=105(2) K and at $T_M(Ru)$=152(1) K. Note, (a) the small peak in the ZFC branch around 125 K. This peak and $T_{irr}$ fit well with the minor and major peaks of the ac susceptibility (Fig. 1). (b) The ferromagnetic-like shape of the FC branches. In general, the ZFC/FC curves do not lend themselves to an easy determination of $T_M(Ru)$, and $T_M(Ru)$ was obtained directly from the temperature dependence of the saturation moment $(M_{sat})$[1]. Here, the non-magnetic Y M/H(T) curve, measured at low H, permits a direct determination of $T_M(Ru)$. At 500 Oe the ZFC and FC branches merge also at $T_{irr}$ (Fig. 2 inset), in contrast to previous measurements on magnetic Ln ions[1], in which $T_{irr}$ is field dependent, and shifted to lower temperatures with H. The irreversibility is washed out for H= 2.5 kOe and both ZFC and FC curves collapse into a single FM-like behavior.

Isothermal M(H) measurements (up to 50 kOe) at various temperatures have been carried out, and the results obtained are exhibited in Figs. 3-4. Below $T_M$, all M(H) curves depend strongly on H (up to 1-5 kOe), until a common slope is reached. The remarkable feature shown in Fig. 3 is the apparent tendency toward saturation at 5 K, but not reaching full saturation even at 50 kOe. Similar behavior was observed in $RuSr_2YCu_2O_8$[21]. This phenomenon is typical of itinerant ferromagnetic materials and reminiscence of the un saturated M(H) curves observed for itinerant ferromagnetic $SrRuO_3$ single crystals at various orientations[22]. The moment at 5 K and 50 kOe is M=$0.71\mu_B$/Ru (Fig. 3). Estimation of the Ru moment at infinite H, by plotting $M^2 \alpha 1/H$ (for high H values), and extrapolating to 1/H=0, yields $M_{max}$= $0.75\mu_B$/Ru, a value which is smaller than $g\mu_B S =1\mu_B$, the expected saturation moment for $Ru^{5+}$ in the low-spin state (g=2 and S=0.5).

The linear part of M(H) in Fig. 3 can be described as: M(H)= $M_{sat}$+ $\chi$H, where $M_{sat}$ =0.63(1)$\mu_B$ is the value obtained by the extrapolation to H=0 and $\chi$=dM/dH is the slope ($8.5*10^{-3}$ emu/mol*Oe =$1.53*10^{-2}\mu_B$/T). This $\chi$ is much larger than the Cu ion contribution to the susceptibility discussed below. Similar M(H) curves have been measured at various temperatures and Fig. 5 shows the temperature dependence of $M_{sat}$. The extrapolated $M_{sat}$ to zero yields $T_M$(Ru)=152(2) K exactly as obtained in Fig. 2. A similar procedure for the magnetic Ln= Dy and Ho ions, yields 150(1) K for both materials, indicating that, within the uncertainty limit, $T_M$(Ru) remains constant regardless of Ln. (The small tail above $T_M$ which appears only for Ln=Y, is probably due to the presence of a small amount of $SrRuO_3$ (not detectable by XRD). The slope $\chi$=M/H obtained at various temperatures is exhibited in Fig. 6. It appears that $\chi$ does not change much up to $T_{irr}$, and then rises up to $T_M$(Ru). Above this temperature the slope follows the Curie-Weiss law discussed below.

At low applied fields, the M(H) curve exhibits a typical ferromagnetic-like hysteresis loop (Fig. 3 inset) similar to that reported in Ref.1-4. Two other characteristic parameters at of the hysteresis loops at 5 K are shown in Fig. 3, namely, the remnant moment, ($M_{rem}$= $0.31\mu_B$/Ru) and the coercive field ($H_C$ = 410 Oe). [For Ln=Dy and Ho at 5 K we obtained: $M_{rem}$ =0.30 and 0.41 $\mu_B$/Ru and $H_C$ = 470 and 320 Oe respectively]. The $M_{rem}$(T) and $H_C$(T) values are exhibited in Fig. 7. Both $M_{rem}$(T) and $H_C$(T) become zero around 100 K which means that essentially no discernible hysteresis is observed at $T_{irr}$(Fig. 4 inset). Surprisingly, at higher temperatures reappearance of the hysteresis loops is obtained (Fig. 4) with a peak at 120 K for $M_{rem}$(T) and $H_C$(T) (Fig. 7 inset) close to the peaks observed in Figs. 1 and 2. In contrast to the FM-like hystersis loop obtained at T< $T_{irr}$ (Fig. 3 inset), the loops above $T_{irr}$ exhibit an AFM like feature. Similar behavior is observed for Ru-1222Dy (Fig. 8) and for Ru-1222Ho[19].

Above $T_M(Ru)$, the $\chi(T)$ curve for $RuSr_2Y_{1.5}Ce_{0.5}Cu_2O_{10}$, measured at 20 kOe up to 400 K, has the typical PM shape and can be fitted by the Curie-Weiss (CW) law: $\chi = \chi_0 + C/(T-\theta)$, where $\chi_0$ is the temperature independent part of $\chi$, C is the Curie constant, and $\theta$ is the CW temperature. In order to isolate the Ru intrinsic magnetic contribution, we also measured (up to 350 K (at 20 kOe)), $\chi(T)$ of $YBa_2Cu_3O_7$ which is roughly temperature independent ($1.8$-$2*10^{-4}$ emu/mol Oe) with values which are two orders of magnitude lower than $\chi(T)$ of Ru-1222Y. After subtracting $2/3\chi(T)$ of $YBa_2Cu_3O_7$ from the measured $\chi(T)$ of Ru-1222Y, we obtained: $\chi_0 = 0.0014$ and C=0.523(1) emu/mol Oe and $\theta = 136(1)$ K, which corresponds to $P_{eff} = 2.05$ $\mu_B$/Ru. [In fact, the same values were obtained for the fit of the raw data]. This $\chi_0$ is comparable to $\chi_0$ of Ru-1212[21], and $\theta = 136$ K is in fair agreement with $T_M(Ru)=152$ K extracted from Figs. 2 and 5. $P_{eff} = 2.05$ $\mu_B$, is somewhat greater than 1.73 $\mu_B$, the expected value of the low-spin state of $Ru^{5+}$ and it is in fair agreement with $P_{eff} = 2.13$ $\mu_B$ obtained in Ref. 4 for $RuSr_2Eu_{1.5}Ce_{0.5}Cu_2O_{10}$, where $P_{eff}$(Ru), was obtained by carefully subtracting all extra contributions to $\chi(T)$. In any case, this $P_{eff}$ is much smaller than 2.8 $\mu_B$/Ru reported for R-1222Eu[23]. [$\chi(T)$ for Ln= Dy and Ho adheres closely to the CW law and the values of C obtained are: 15.6 and 12.1 emu/mol Oe respectively. These values are much smaller than the theoretical values expected for $Dy^{3+}$ and $Ho^{3+}$ free (1.73 $\mu_B$) ions, probably due to a strong crystal field effects. However, this is of a little interest in the present discussion.]

**Discussion**

In a recent paper[13] we have shown that a small amount (2.5 at%) of Zn suppresses completely the SC state in $RuSr_2Eu_{2-x}Ce_x(Cu_{1-x}Zn_x)_2O_{10}$ and that the magnetic properties of the Ru sublattice are not effected by the presence or the absence of this SC state. Our general picture is that (a) in the Ru-1222 system, all compounds have a similar magnetic structure and (b) that the two states are practically decoupled. Therefore, the study of non-SC and non-magnetic Y ions permits an easier interpretation of the intrinsic Ru-O sub-lattice magnetism, in particular that of the PM state.

**The qualitative magnetic structure of $RuSr_2Y_{1.5}Ce_{0.5}Cu_2O_{10}$.** While our data described here do not include any determination of the magnetic structure of the Ru sublattice, the results are compatible with a simple model, which is, however, of use in understanding the qualitative features at low applied fields. Starting from high to low temperatures, the magnetic behavior is basically divided into 4 regions.

(i) At elevated temperatures, the paramagnetic net Ru moment is well described by the CW law, and the extracted values for Ru-1222Y are: $P_{eff}$ =2.05 $\mu_B$ and $\theta$=136 K. this $P_{eff}$ is somewhat greater than 1.73 $\mu_B$, the calculated value of $Ru^{5+}$ in the low-spin state and suggests that the assumption of completely localized moments is not adequate for this system.

(ii) At $T_M$ (=152 K), the Ru sub-lattice becomes basically AFM ordered at low applied fields. This interpretation is supported, by the small peaks observed in both: the ac susceptibility (Fig. 1) and in the ZFC branch when measured at low fields (Fig.2). At higher applied fields a metamagnetic transition is induced. In general, metamagnetism in insulating AFM systems refers to the magnetic transition produced by an external H, when the strength of the field equals or exceeds the exchange coupling between the magnetic moments. In Ru-1222Y, the AFM Ru moments are realigned through a spin-flip process by H, to form the AFM-like shape hysteresis loops observed in Figs 4 and 8. Note the difference between the AFM-like and the FM-like (Fig. 3) hysteresis loops.

(iii) At $T_{irr}$=100 K a weak ferromagnetism is induced, which originates from canting of the Ru moments. $T_{irr}$ is defined as the merging point of the low field ZFC and FC branches, or alternatively, at the temperature in which $M_{rem}$ and $H_C$ first disappear. This canting (which is significantly different from the spin-flip mechanism in (ii)) arises from the DM anti-symmetric super-exchange interaction, which by symmetry, follows from the fact that the $RuO_6$ octahedra tilt away from the crystallographic c axis[17]. The ratio of $M_{rem}$ /$M_{max}$ (0.31/ 0.75=0.41) is also consistent with W-FM order in this system. The nature of the local structural distortions causing DM exchange coupling is not presently known.

The exact structure of the Ru spin ordering in the third region is still debated and no conclusions have been reached yet. Neutron diffraction measurements are required to precisely determine the nature of the magnetic order. Three scenarios can be described. (I) Our preferred one, is that a net long-range W-FM component exists, which in principle arises from the canting procedure mentioned above. The saturation moment of $Ru^{5+}$ in the low spin state is 1 $\mu_B$/Ru (3 $\mu_B$/Ru for $Ru^{5+}$ in the high spin state). We note that 0.75$\mu_B$/Ru obtained, is a large fraction of 1 $\mu_B$, implying a very large canting angle (49°) of the AFM Ru moments. (II) Alternatively, a frequency dependence cusp was observed in the ac susceptibility of Ru-1222, which is interpreted as a spin glass behavior[24]. A spin glass is a collection of magnetic moments whose state is frozen and disordered. To produce such a state, two ingredients are necessary: frustration and partial randomness of interaction between the magnetic moments. Although Ru ions in Ru-1222 are arranged in a strictly periodic order, it is possible that the incomplete oxygen stoichiometry and/or the small distortion of the oxygen octahedra are involved in inducing frustration, which, in turn leads to glassy behavior. (III) The un-saturated magnetization curves at low temperatures may suggest (similar to $SrRuO_3$) itinerant-electron magnetism in this system. Consequently, we used the conventional so-called Arrott-plots to determine the magnetic transition. In Fig. 5 (inset) we have plotted the field values of Ru-1222Y at various temperatures and the magnetic transition extracted is 125 (3) K, midway between $T_{irr}$ and $T_M$. A supporting evidences to this scenario are: (a) the high $P_{eff}/M_{sat} \sim 3$ ratio and (b) the $P_{eff} = 2.05$ $\mu_B$ value which exceeds the expected value for a localized $Ru^{5+}$ low spin state.

(iv) For Ln= Eu and Gd, SC is induced at $T_C$. $T_C < T_M$ depends strongly on Ce (as hole carriers) and on oxygen concentrations[1-5]. Below $T_C$, both SC and weak-ferromagnetic states coexist and the two states are practically decoupled[13].

However, the present three scenarios, differ completely from the magnetic phase separation of AFM and FM nano-domain particle model[3] which assumes that in Ru-1222, minor part of the material becomes FM at $T_M$ and persists down to low temperatures, whereas the major part orders AFM (at the so-called $T_{irr}$) then becomes SC at $T_C$. In this scenario, the unusual SC state is well understood. However this model cannot be reconciled with the accumulated data presented here: (a) The high 0.75 $\mu_B$/Ru moment at 5 K (b) The continuous $M_{sat}$ curve (Fig. 5) which does not show any inflection at $T_{irr}$ or at lower temperatures (c) The reopening of hysteresis loops above 100 K, and as a result the increase of both $M_{rem}$ and $H_C$ (Fig. 7). According to this model, the hysteresis loops opened at $T_M$ would remain all the way down to low temperatures and both $M_{rem}$ and $H_C$ would increase continuously, or at least remain constant. (d) The different shape of the hysteresis loops observed below and above $T_{irr}$.

As a final point of interest we may compare the whole magnetic behavior of Ru-1222Y to that of ferromagnetic $La(Fe_xSi_{1-x})_{13}$ intermetallic compounds[25]. In both systems above the FM (or W-FM in Ru-1222) transition, small hysteresis loops are opened and both systems exhibit meta-magnetic transitions. In $La(Fe_xSi_{1-x})_{13}$, the itinerant electron metamagnetic transition observed above $T_C$, is a result of thermal activations that make the FM energy minimum shallower than the PM one, resulting in the PM state above $T_C$. By applying a field, the FM energy minimum again becomes lower than the paramagnetic one and a metamagnetic transition from PM to the itinerant FM state is induced. As a result small hysteresis loops are observed above $T_C$. The only one difference is that in the Ru-1222 system, the ground state in region (ii) is AFM and the external H induces a spin-flip transition.

**In conclusion**, the magnetic behavior of all non-SC materials studied is practically the same, and the magnetic parameters, such as $T_{irr}$ $T_M$ $H_C$ and $M_{sat}$ are similar. Two steps in the magnetic behavior are presented. At $T_M$ =152(2) K all materials become AFM ordered and a metamagnetic transition is induced by the applied field. and typical AFM hysteresis loops are observed.  At $T_{irr}$ ranging from 90-100 K, a W-FM state is induced, originating from canting of Ru moments caused from the DM anti-symmetric super-exchange interaction, and follows from the fact that the $RuO_6$ octahedra tilt away from the crystallographic c axis. The PM parameters extracted indicate that Ru is pentavalent in the low spin state. The saturation moment at 5 K is 75% of the expected value for $Ru^{5+}$ indicating a W-FM state at low temperatures.

**Acknowledgments** We are grateful to E. Galstyan for assistance in the ac measurements. This research was supported by the Israel Academy of Science and Technology,  and by the Klachky Foundation for Superconductivity.

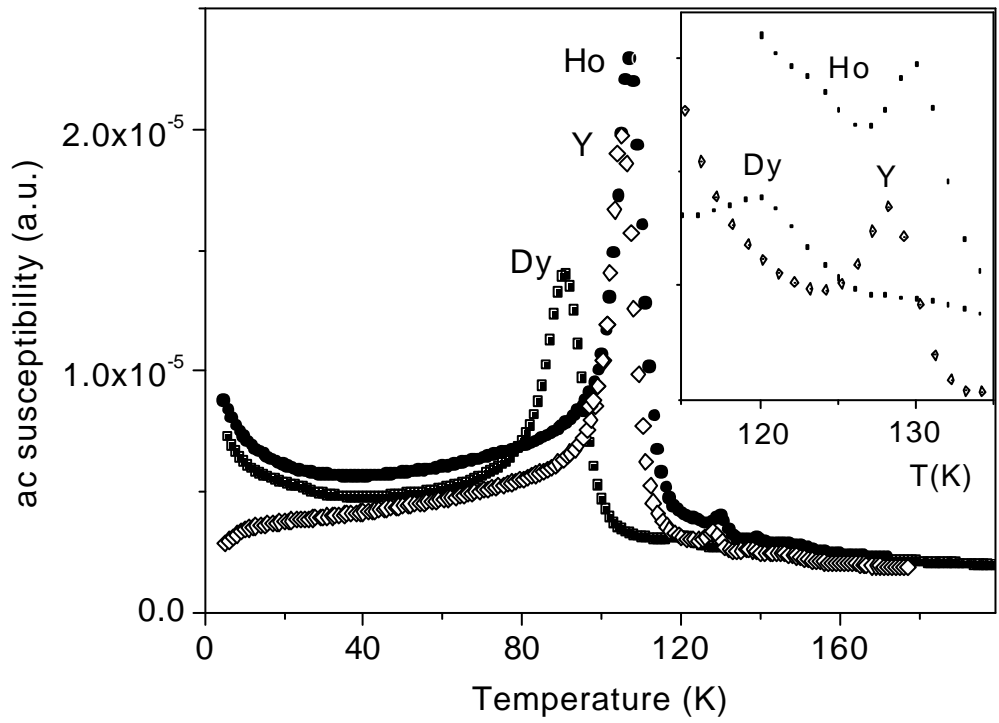

Fig 1 The real part of the ac susceptibility of $RuSr_2Ln_{1.5}Ce_{0.5}Cu_2O_{10}$

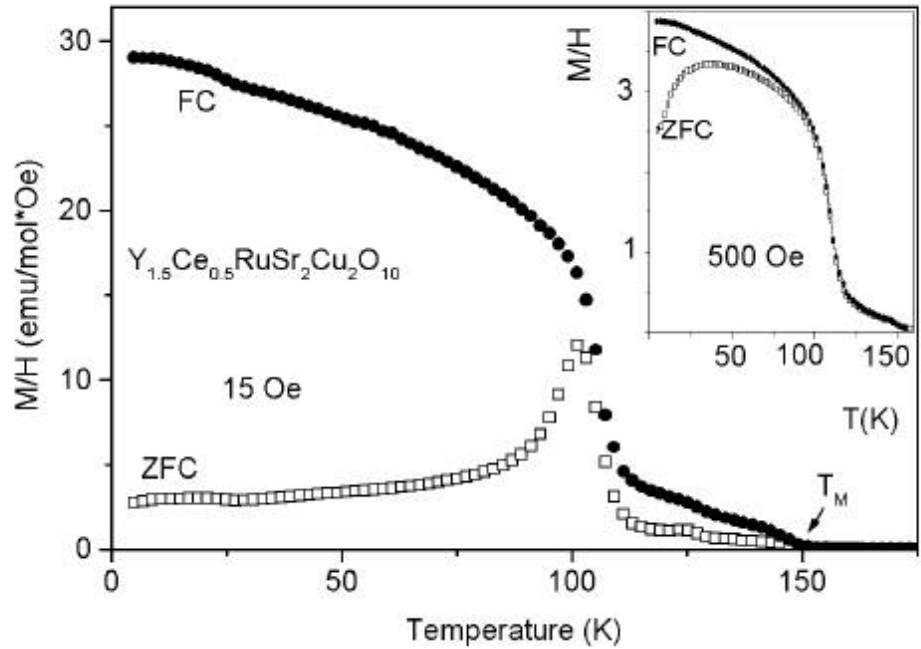

Fig 2 ZFC and FC M/H for Ru-1222Y measured at 15 Oe and 500 Oe (inset).

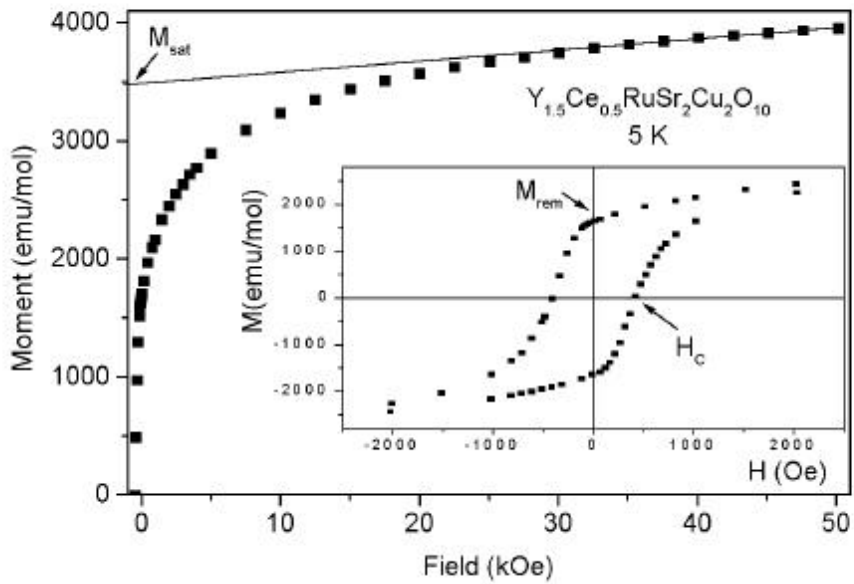

Fig 3 The isothermal magnetization at 5 K and the ferromagnetic-like hysteresis loop (inset)

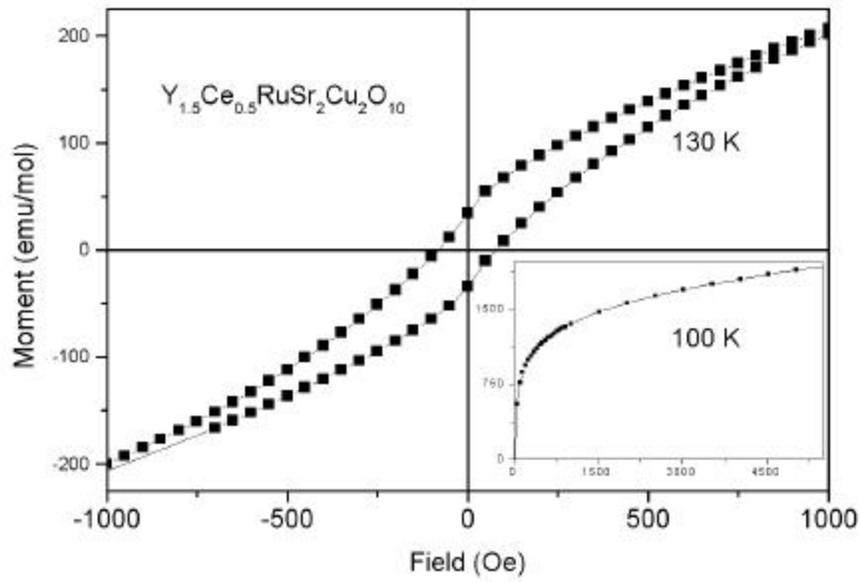

Fig. 4 The hysteresis loop opens at 130 K and the magnetization at 100 K(inset)

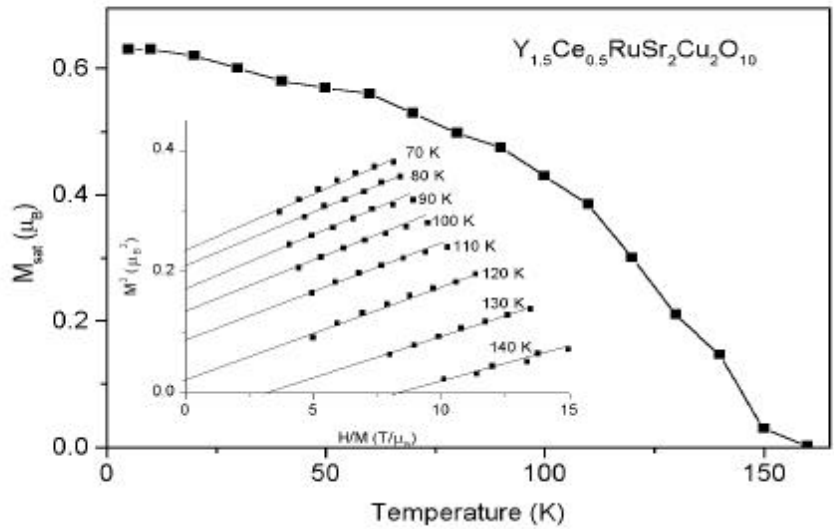

Fig. 5 The temperature dependence of the saturation moment of R-1222Y. The Arrott plots are shown in the inset

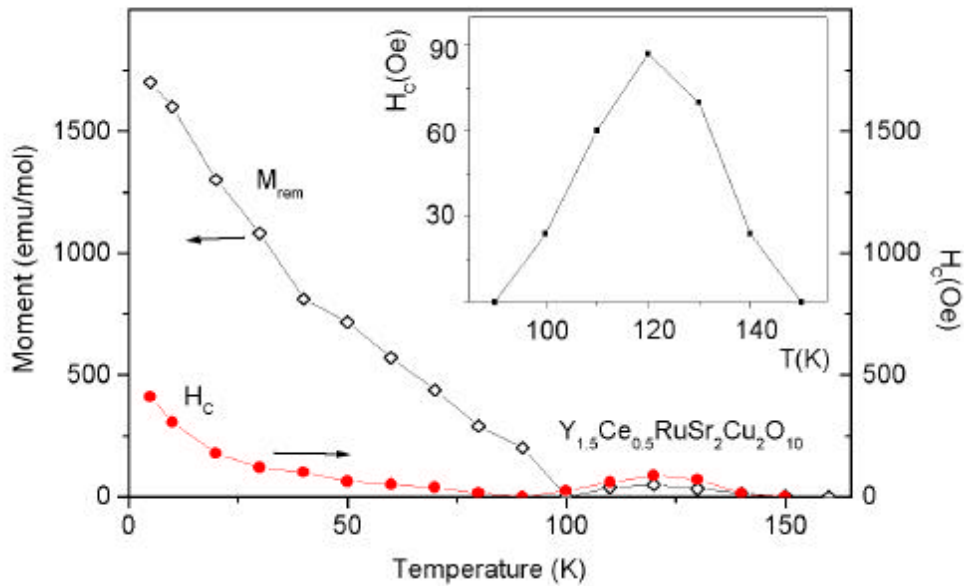

Fig. 7 The temperature dependence of the remanent and coercive field of R-1222Y

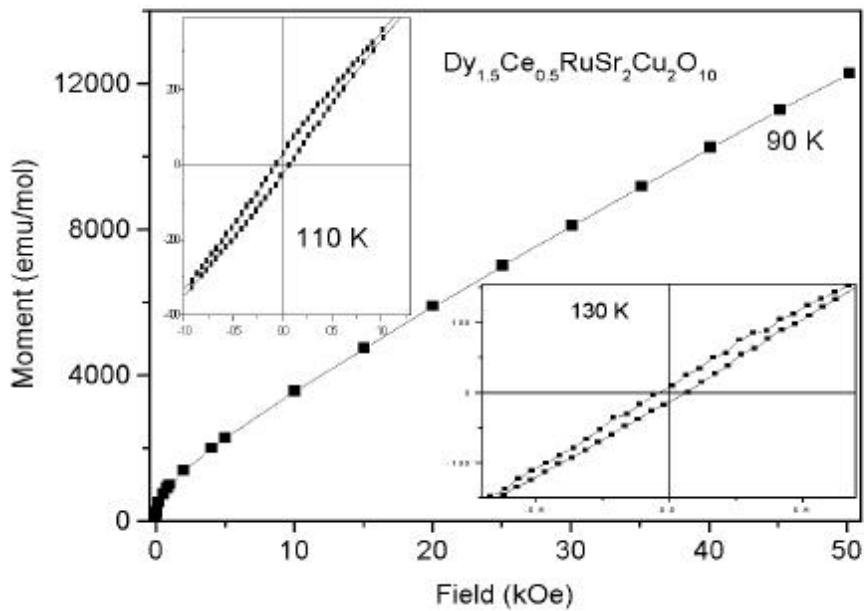

Fig. 8 The isothermal magnetization curves for R-1222Dy at 90, 110 and 130 K. Note the absence of hysteresis at 90 K.

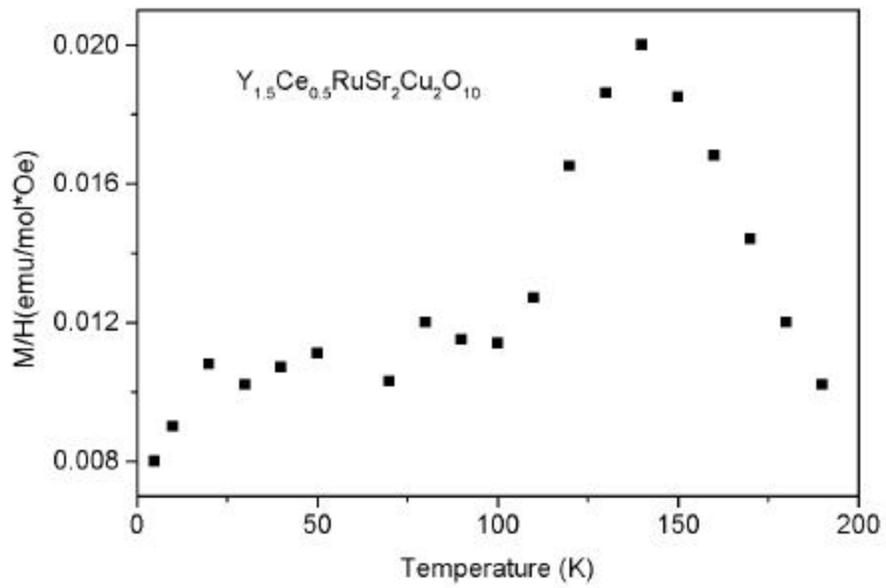

Fig. 6. The slope of the high field values, at various temperatures.